\documentclass[a4paper]{article}

\usepackage{multirow}
\usepackage{INTERSPEECH2019}

\title{End-to-End ASR for Code-switched Hindi-English Speech}
\name{Brij Mohan Lal Srivastava$^1$\sthanks{Work done during internship at Microsoft Research India}, Basil Abraham$^2$, \\ Sunayana Sitaram$^2$, Rupesh Mehta$^2$, Preethi Jyothi$^3$}
\address{
  $^1$INRIA, France\\
  $^2$Microsoft Corporation, India, $^3$IIT Bombay, India}
\email{brij.srivastava@inria.fr, \{basil.abraham, sunayana.sitaram, rupesh.mehta\}@microsoft.com, pjyothi@cse.iitb.ac.in}

\begin{document}
%
\maketitle
\begin{abstract}
End-to-end (E2E) models have been explored for large speech corpora and have been found to match or outperform traditional pipeline-based systems in some languages. However, most prior work on end-to-end models use speech corpora exceeding hundreds or thousands of hours. In this study, we explore end-to-end models for code-switched Hindi-English language with less than 50 hours of data. We utilize two specific measures to improve network performance in the low-resource setting, namely multi-task learning (MTL) and balancing the corpus to deal with the inherent {\it class imbalance} problem i.e.~the skewed frequency distribution over graphemes. We compare the results of the proposed approaches with traditional, cascaded ASR systems. While the lack of data adversely affects the performance of end-to-end models, we see promising improvements with MTL and balancing the corpus.
\end{abstract}


\section{Introduction}
\label{sec:intro}

Unlike the traditional automatic speech recognition (ASR) pipeline where several independently optimized modules are integrated (typically using weighted finite state transducers), end-to-end (E2E) architectures instead optimize a single neural network that maps acoustic events to grapheme sequences. Recently, there has been a substantial amount of work on end-to-end architectures with several studies conducted on huge datasets containing several hundreds or thousands of hours of speech data~\cite{weng2018improving,zeghidour2018end,zeyer2018improved,toshniwal2017multilingual}. In this paper, we summarize our challenges and results when trying to optimize two types of popular end-to-end architectures, namely connectionist temporal classification (CTC)~\cite{graves2014towards} and attention-based models~\cite{chan2016listen} for low-resource code-switched Hindi-English ASR.


Firstly, we compare the performance of CTC and attention-based architectures with time delay neural network (TDNN) based acoustic models. We then combine both CTC and attention-based loss functions in a multi-task setting~\cite{kim2017joint} and observe the change in performance as we vary the weights given to each loss function. 




\begin{table*}[h]
\caption{\textit{Data description}} 
\centering
\begin{tabular}{|c|c|c|c|c|c|}
  \hline
 Language & Train (hrs) & Dev (hrs) & Test (hrs) & Graphemes & Phonemes \\
 \hline
 Hindi-English CS & 46.65 & 5.83 & 5.70 & 101 & 89\\
  \hline
\end{tabular}
\label{tab:datadesc}
\end{table*}

Second, we propose an approach to handle the low-resource setting based on the observation that code-switched data inherently exhibits class-imbalance issue. The sample distribution over graphemes is highly skewed by nature due to their frequency of usage. This skewed distribution does not affect high-resource languages as much due to sufficient availability of samples per class. However, in the code-switched setting which is low-resource due to the lack of annotated data, this imbalance poses a significant challenge to optimize the neural network. We apply oversampling-based techniques to address this issue and investigate its benefits.

Section~\ref{sec:prior} describes relevant prior work on end-to-end systems for code-switched speech recognition.  Section~\ref{sec:approaches} explains the approaches we explore. Section~\ref{sec:expt} lists the experiments conducted to validate our proposed techniques. Section~\ref{sec:conc} concludes the paper and describes future directions.

\section{Relation to prior work}
\label{sec:prior}
Graves et al.~\cite{graves2006connectionist} first proposed the CTC loss function to predict the underlying sequence of phonemes in a speech signal without using any a priori phone alignment information. They achieved around 30\% label error rate on the TIMIT corpus. Graves et al.~\cite{graves2014towards} proposed an end-to-end model that yielded a 27\% WER on the Wall Street Journal (WSJ) corpus, without the use of a lexicon and language model (LM), and the WER reduced to 6.7\% when a trigram LM was used. Miao et al.~\cite{miao2015eesen} published a toolkit, EESEN, which used RNNs and WFST-based decoding for end-to-end speech recognition. Their experiments with the WSJ corpus yielded around 8\% WER when a trigram LM was applied.


Convolutional neural network (CNN) based architectures have also been employed within end-to-end ASR~\cite{collobert:2016,liptchinsky:2017,zhang2017very,zhang2017towards} and they achieve WERs in the range of 7-10\% on TIMIT, WSJ and the Librispeech corpus (when augmented with a lexicon and LM). Zhang et al.~\cite{zhang2017very} achieve around 10\% WER on the WSJ task without any lexicon or LM using a very deep (15 layers) convolutional architecture.

Attention-based models have also been explored in end-to-end speech recognition~\cite{chorowski2014end,chorowski2015attention,bahdanau2016end,kim2017joint,chan2016listen}. These models achieve a WER of 18\% without an LM on WSJ, which is further cut down by half by using a trigram LM. The \emph{Listen, Attend and Spell (LAS)} model~\cite{chan2016listen} is currently state-of-the-art and there are several ongoing studies to improve it for various settings such as multilingual ~\cite{toshniwal2017multilingual} and multi-dialect~\cite{li2017multi} speech recognition. These systems are trained over 12000-15000 hours of multilingual and multi-dialect speech data and produce WERs within the range of 17-21\% without the use of an LM. Chiu et al.~\cite{chiu2017state} also proposed techniques like word-piece models, multi-headed attention, minimum WER training and second-pass LM rescoring to further improve LAS models and obtained a 4\% WER on a 12000-hour voice search task.

For most of the systems described above, the training data size ranges from 200-12000 hours, and the language is predominantly English except~\cite{toshniwal2017multilingual}, who present their work on nine Indian languages and their training data size is around 1500 hours. We build our systems in far more limited settings with around 50 hours of data.


Recently, two papers have explored end-to-end ASR for code-switching. Winata et al.~\cite{winata2018towards} proposed training a CTC based model on 100 hours of Mandarin-English code-switched speech. They pretrain the model alternately using monolingual English and Mandarin speech and then jointly train on a code-switched corpus. Luo et al.~\cite{luo2018towards} is similar to our current work; they explore a hybrid CTC/attention end-to-end model to recognize Mandarin-English speech and study the effect of larger units for decoding. They also study the inclusion of language identification within the ASR framework.








\section{Our approach}
\label{sec:approaches}

In this section, we describe our two main approaches for the low-resource code-switched setting.





\begin{figure*}[htb]
  \centering
  \includegraphics[width=\linewidth]{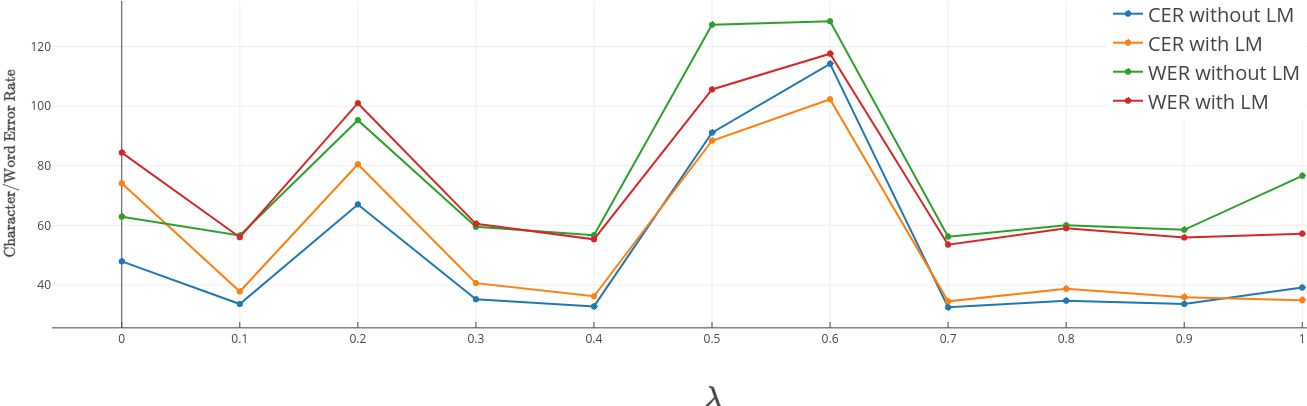}
\caption{Character/word error rates for code-switched Hindi-English specific to different $\lambda$ values.}
\label{fig:mtl-range}
\end{figure*}

\subsection{Multi-task learning}

We make use of the multi-task learning framework outlined by~\cite{kim2017joint}. In this approach, we jointly optimize a convex combination of two loss functions: \emph{CTC} and \emph{attention}. \emph{CTC loss} is  defined as the negative log likelihood of the true label sequence ($y^*$) given the acoustic features ($x$).

\begin{equation}
L_{CTC} = -\log P(y^* | x)
\label{ctcloss}
\end{equation}

In practise, CTC assumes a set of all possible label sequences $\Phi(y)$ and computes $P(y^*|x)$ as the marginal probability over $\Phi(y)$ using dynamic programming. While computing $P(\phi|x)$, CTC assumes independence between labels $l_i$.

\begin{equation}
    P(y^*|x) = \sum_{\phi \in \Phi(y)} P(\phi|x)
\end{equation}
\begin{equation}
    P(\phi|x) = \prod_{i=1}^T P(l_i | x)
\end{equation}


\emph{Attention loss} is defined as the sum of negative log likelihood of true phoneme prediction at each time step $t$ conditioned over the previous $t-1$ phoneme predictions and the decoder state ($x$) which represents the acoustic features attended for the current time step. It is computed using:

\begin{equation}
L_{ATT} = - \sum_t \log P(y^*_t | x, y^*_{1:t-1})
\end{equation}

We observe that the attention loss converges faster but is inferior in performance to CTC in our low-resource setting. We empirically investigate this issue further and find that combining both CTC and attention loss functions remedy this problem to an extent. However, this combination must be implemented carefully as explained in Section~\ref{sec:expt}.

The two loss functions are combined as follows:

\begin{equation}
L_{MTL} = \lambda * L_{CTC} + (1-\lambda) * L_{ATT}
\end{equation}

Here $\lambda$ is the convex combination hyperparameter and we experiment with different values of $\lambda$ to ascertain the role of both the loss functions in low-resource settings (refer to Section~\ref{mtl-expt-details}).

\subsection{Class imbalance}
\label{sec:imb-explain}

\begin{figure}[h]
  \centering
  \includegraphics[width=\linewidth]{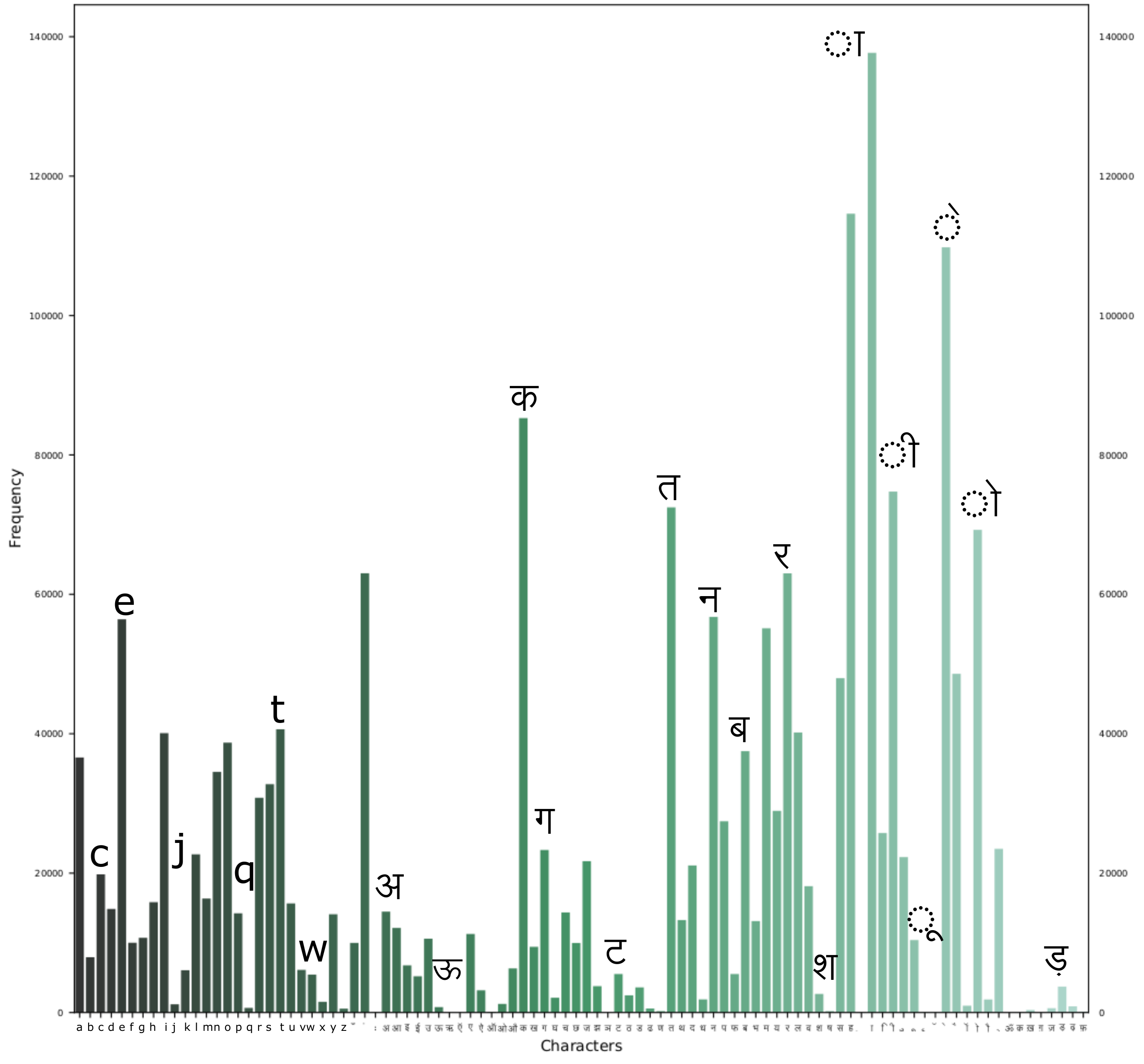}
\caption{Training set character distribution for code-switched Hindi-English. The y-axis plots the frequency of the corresponding character on the x-axis. Some prominent bars are annotated with the characters they represent.}
\label{fig:char-dist-proof}
\end{figure}

There have been several efforts to remedy the issue of class imbalance in machine learning algorithms. This is a critical issue for end-to-end models since natural language is inherently imbalanced in terms of character classes and the models will not learn reliable estimates for classes that have very low coverage in the dataset. In prior work, three dominant approaches have been used to address this issue~\cite{zhou2006training,buda2017systematic}: random minority oversampling, random majority undersampling and cost-effective learning. In this study, we explore oversampling since undersampling is not feasible for the low-resource scenario and we plan to explore cost-effective learning as future work.


To oversample the training dataset, we compute the character frequency distribution of the corpus (refer to Figure~\ref{fig:char-dist-proof}) and for each sentence. We identify the sentences which contain rare characters and simply repeat them to match the mean frequency. We define rare characters as ones having lower frequency than the mean frequency. We try to avoid the duplication of sentences which contain a lot of frequent characters but it is unavoidable due to their pervasiveness.

\section{Experiments}
\label{sec:expt}

\subsection{Data}

Table \ref{tab:datadesc} describes the conversational Hindi-English data used in our study; more details can be found in~ \cite{sivasankaran2018phone}. The grapheme set includes $\langle space\rangle$, $\langle sos\rangle$, $\langle eos\rangle$ and $\langle unk\rangle$ identifiers for spaces, start-of-sentence, end-of-sentence and unknown characters.

\begin{table*}[t]
\centering
\caption{Character/Word Error Rates (CER/WER) of standard ASR systems vs. E2E ASR systems}
\label{tab:all-wer}
\begin{tabular}{|c|ccc|cccccc|}
\hline
\textbf{\begin{tabular}[c]{@{}c@{}}Language/Model\end{tabular}} & \multicolumn{1}{c|}{GMM-HMM} & \multicolumn{1}{c|}{DNN} & TDNN           & \multicolumn{2}{c|}{CTC}                            & \multicolumn{2}{c|}{LAS}                            & \multicolumn{2}{c|}{MTL}                 \\ \cline{2-10} 
                                                                                   & \multicolumn{1}{c|}{WER}     & \multicolumn{1}{c|}{WER} & WER            & \multicolumn{1}{c|}{CER} & \multicolumn{1}{c|}{WER} & \multicolumn{1}{c|}{CER} & \multicolumn{1}{c|}{WER} & \multicolumn{1}{c|}{CER} & WER           \\ \hline
Hindi-English CS                                                                   & 40.21                        & 33.78                    & \textbf{31.78} &  34.9                        & 57.2                     &  47.9                        & 62.9                     &  34.1                        & \textbf{52.3} \\ \hline
\end{tabular}
\end{table*}


\begin{figure*}[htb]

\begin{minipage}[b]{0.33\linewidth}
  \centering
  \centerline{\includegraphics[width=\linewidth]{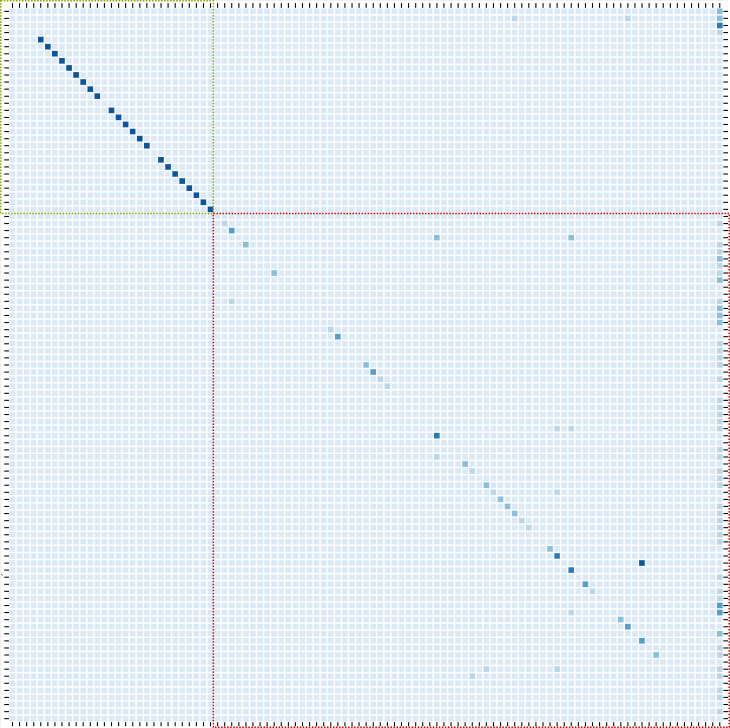}}
  \centerline{(a) $\lambda = 0.2$}\medskip
\end{minipage}
\begin{minipage}[b]{.33\linewidth}
  \centering
  \centerline{\includegraphics[width=\linewidth]{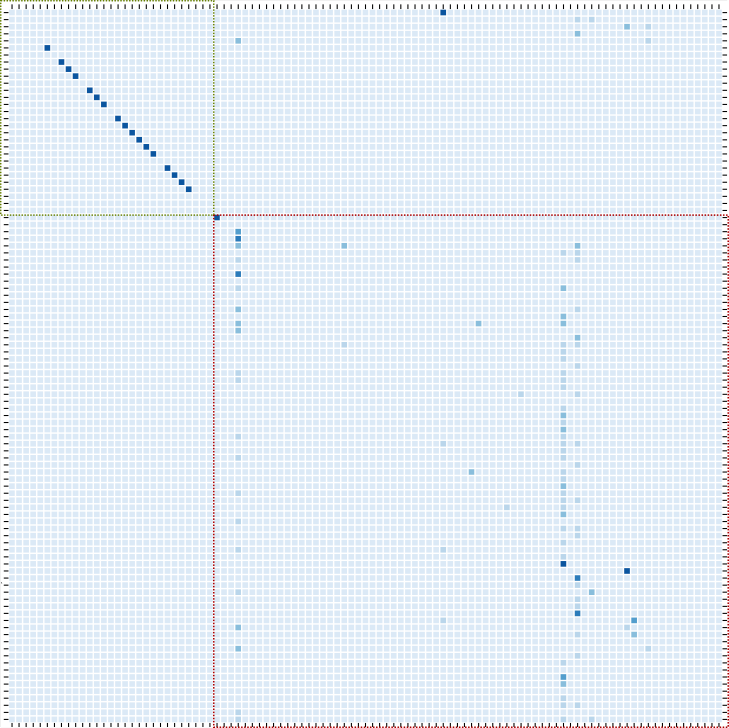}}
  \centerline{(b) $\lambda = 0.5$}\medskip
\end{minipage}
\begin{minipage}[b]{0.33\linewidth}
  \centering
  \centerline{\includegraphics[width=\linewidth]{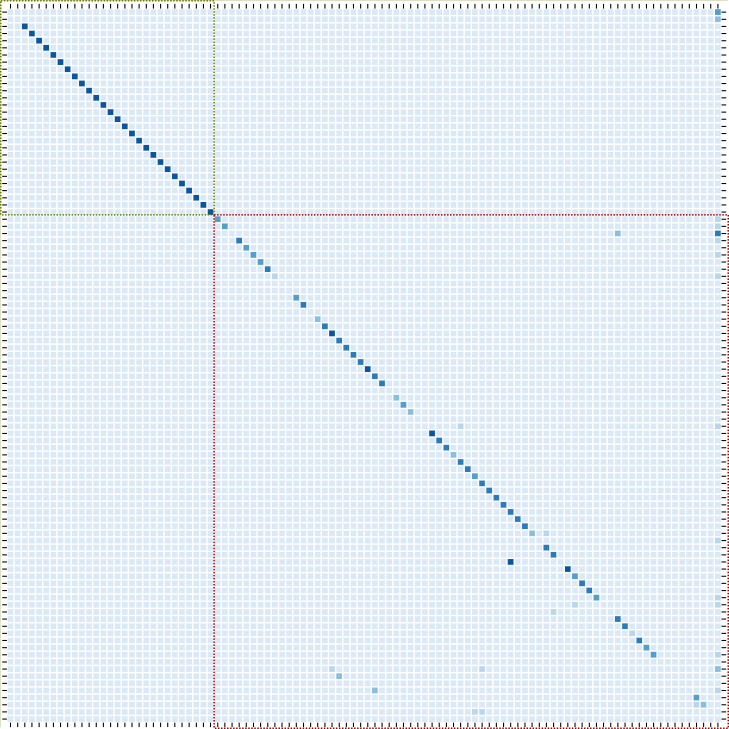}}
  \centerline{(c) $\lambda = 0.7$}\medskip
\end{minipage}
\caption{Character confusion matrices for Hindi-English CS data and different MTL $\lambda$ values. The dotted green region indicates confusion for Hindi characters and the dotted red region shows confusion for English characters.}
\label{fig:mtl-confusion}
\end{figure*}



\subsection{Baselines}

We compare our approaches against baselines built using the traditional ASR pipeline. We use the Kaldi~\cite{povey2011kaldi} toolkit to built GMM-HMM, DNN and TDNN models using \texttt{wsj/s5} scripts. TDNN performs the best as shown in Table~\ref{tab:all-wer}.

All the systems are trained over two 6-core Intel Xeon machines with a single Tesla P100 GPU. It took 20-30 hours to train each system. The training dataset size for class imbalance experiment was approximately 20 times the original, hence it took around 350 hours to complete the experiment. 

We extract 40-dimensional melscale filterbank coefficients with their first and second order derivatives to train GMM-HMM systems. Further, to train 5-layered p-norm DNN systems, we obtain feature-space maximum likelihood linear regression (fMLLR) which are appended with 100-dimensional iVectors to train 6-layered TDNN systems. The system is trained over the phone set mentioned in Table~\ref{tab:datadesc}.








\subsection{MTL experiments}
\label{mtl-expt-details}




We combine the two end-to-end approaches for speech recognition within the MTL framework. We also create models with just CTC and Listen-Attend-Spell (LAS)~\cite{chan2016listen} architectures for comparison.



LAS is implemented using ESPnet~\cite{watanabe2018espnet} framework and modified to suit the low-resource data requirements. The encoder contains 4 layers of stacked Bidirectional LSTMs with a pyramid structure for sub-sampling. We experiment with three different types of attention: vanilla, location aware and windowed. We also regularize the network by adding Gaussian noise of standard deviation 0.01 to the input and a dropout rate of 0.5 over the output of encoder. The decoder is composed of a two layer LSTM network with each layer in the encoder having 320 units and the decoder having 300 units. To make the output of decoder robust to noisy predictions, we sample from the output during training with the probability 0.1. The network uses a shared encoder and two separate decoders for CTC and Attention.

We decode each test utterance using a beam search decoder with 50 beams and finally obtain 50 n-best character sequences per utterance. Each network is trained for 15 epochs to minimize the convex combination ($\lambda$) of CTC and Attention loss using Adadelta~\cite{zeiler2012adadelta} optimizer to predict the target sequence with an $\langle eos\rangle$ token added. All the models converge within the range of 15-21k gradient steps. Table~\ref{tab:all-wer} compares the results between the baseline systems and the E2E architectures. MTL results indicate the WER for the best $\lambda$. We observe that MTL outperforms CTC and LAS models.


We experiment with the following range $0.0 \le \lambda \le 1.0$ and observe that all the systems perform well only when the value of $\lambda$ is close to either 0.0 or 1.0. The word/character error rates suddenly peak when the value is close to 0.5 as shown in Figure~\ref{fig:mtl-range}. Robust results were obtained within the range $0.7 \le \lambda \le 0.9$ for all the languages. This might be due to the fact that the attention model is too flexible to allow non-sequential alignments for the output of the decoder, which is non-permissible for speech recognition as described by~\cite{watanabe2017hybrid}. CTC acts as a regularizer due to its Markov chain assumption and prunes the alignments which are non-sequential. Hence, giving higher weight to CTC (higher $\lambda$) might produce robust results as compared to attention.

We conduct the MTL range experiment for various values of $\lambda$ as shown in Figure~\ref{fig:mtl-range}. The x-axis ranges from 0 to 1 for the values of $\lambda$ with step size of 0.1. The y-axis indicates the character or word error rate (CER/WER) which typically ranges from 0 to 100, but may exceed the bound of 100 if the hypothesis is much larger than the reference transcription. We plot the CER/WER with and without rescoring the predicted hypotheses with a character-level language model as mentioned in the legend. We observe that values closer to 0.5 produce unstable results exceeding the WER more than 100. We also plot character-level confusion matrices (refer to Figure~\ref{fig:mtl-confusion}) for 3 MTL values (0.2, 0.5, 0.7). These are $102 \times 102$ matrices for $101$ code-switched Hindi-English characters and an $\epsilon$ (right and bottom-most) symbol for representing insertions/deletions. A better model would have a prominent diagonal. We clearly observe that in case of $\lambda = 0.2$, English characters (green region) are recognized well but Hindi (red region) are confused. In case of $\lambda = 0.5$ the character confusion increases dramatically for both English and Hindi. $\lambda = 0.7$, in comparison, performs very well. Also note in Figure~\ref{fig:mtl-range} that $0.7 \le \lambda \le 0.9$ produces stable results. Hence, we conclude this range is optimal for our low-resource setting.

\subsection{Class imbalance experiments}

We oversampled the Hindi-English CS training dataset based on the approach described in Section~\ref{sec:imb-explain}. Table~\ref{tab:imb-stats} contains the statistics on the overall character frequencies in our training corpus. 

\begin{table}[bh!]
\caption{Character frequencies for the training corpus before and after applying oversampling}
\begin{tabular}{c|c|c|c|c|}
\cline{2-5}
                                      & \textbf{Min} & \textbf{Max} & \textbf{Avg.} & \textbf{Median} \\ \hline
\multicolumn{1}{|c|}{\textbf{Before}} & 3            & 137695       & 20546.13      & 10495.0         \\ \hline
\multicolumn{1}{|c|}{\textbf{After}}  & 20631        & 6430338      & 897325.04     & 472576.0        \\ \hline
\end{tabular}
\label{tab:imb-stats}
\end{table}

Note that the minimum character frequency is 3 and the maximum is over 0.1 million which is a huge imbalance as also visible from Figure~\ref{fig:char-dist-proof}. We oversample the characters which are below the mean frequency which brings the minimum frequency to be around 20k. The augmented dataset size increases to 1.2M sentences as compared to 41K sentences in the original dataset. We observe that the architecture used in Section~\ref{mtl-expt-details} with $\lambda = 0.7$ converges to a lower loss in fewer epochs with the balanced dataset than with the imbalanced one, although the time required for each epoch is significantly higher. The test set WER performance also improves by 5\% absolute producing a WER of 47.2\%.

\section{Conclusion \& Future directions}
\label{sec:conc}





In this work, we investigate two approaches for end-to-end ASR of   code-switched speech in a low-resource setting. We explore multi-task learning by combining CTC and attention loss and notice that there is a certain range of the combination parameter $\lambda$ which produces robust performance. We also notice that most of the errors made by our models are substitution errors between graphemes which are typically confused (e.g. nasal consonants with nasal diacritics). We plan to alleviate this issue by building more robust language models which can help disambiguate between such instances. 

Next, we observe that there is an inherent class imbalance problem in end-to-end speech recognition for low-resource languages and propose an oversampling based solution which improves the performance of our best code-switched model by 5\% absolute. In future work, we will explore gradient-based approaches to deal with the class imbalance challenge. We plan to implement intermediate layers which would scale the gradients in proportion to the fraction of class samples in a mini-batch.

\bibliographystyle{IEEEtran}
\bibliography{strings,refs}

\end{document}